\documentclass[aps,prl,preprint]{revtex4}

\usepackage{epsfig}

\begin{document}
\author{Yu Jiang$^{1}$, Hui-xia Zhu$^{1,2}$, Wei-min Sun$^{1,3}$ and Hong-shi Zong$^{1,3}$}
\address{$^{1}$ Department of Physics, Nanjing University, Nanjing 210093, China}
\address{$^{2}$ Department of Physics, Anhui Normal University, Wuhu, 241000, China}
\address{$^{3}$ Joint Center for Particle, Nuclear Physics and Cosmology, Nanjing 210093, China}
\title{The Quark Number Susceptibility in
Hard-Thermal-Loop Approximation}

\begin{abstract}
With the aid of the vector Ward-Takahashi identity we derive a
general formula for the quark number susceptibility (QNS) which
expresses the QNS as an integral expression only involving the full
quark propagator at finite temperature and chemical potential. The
QNS at finite temperature and zero chemical potential is calculated
with the dressed quark propagator in the Hard-Thermal-Loop (HTL)
approximation. A comparison of our result with the results of QNS in
HTL approximation in previous literatures is given.
\bigskip


\bigskip

E-mail: zonghs@chenwang.nju.edu.cn.

\bigskip

PACS Number(s): 12.38 Mh, 11.10 Wx

\end{abstract}

\maketitle

Analysis of fluctuations is a powerful method for characterizing the
thermodynamical properties of a system. It is believed that enhanced
fluctuations are an essential characteristic of phase transitions
\cite{Kunihiro,Stephanov,Brown,Hatta,Fujii,Son}. A measure of the
intrinsic statistical fluctuations in a system close to thermal
equilibrium is provided by the corresponding susceptibilities. Among
various QCD susceptibilities, which are associated with the response
of the QCD vacuum to external sources, the quark number
susceptibility (QNS) plays an important role in identifying the
critical end point in the QCD phase diagram
\cite{Stephanov,Hatta1,Foder,Stephanov1,Lungwitz,Schaefer,He}. To
date, there has been a great amount of related works for calculating the QNS, for example,
lattice QCD simulation \cite{Elia,Allton,Gottlieb,Gavai}, the
Nambu-Jona-Lasinio model \cite{Fujii,Kunihiro,Sasaki,Ratti} and
Polyakov-Nambu-Jona-Lasinio model (PNJL) \cite{Ghosh1,Ghosh2}, the
QCD Dyson-Schwinger equation approach \cite{He,He1,He2,He3}, the
hard-thermal-loop (HTL) approximation
\cite{Blaizot,Chakraborty,Blaizot1,Chakraborty1}, and so on. In this
paper we try to revisit the calculation of the QNS in the HTL
approximation by a direct method.

The partition function of QCD at zero temperature and finite
chemical potential reads
\begin{eqnarray}
{\cal{Z}}[\mu]&=&\int{\cal{D}}\bar{q}{\cal{D}}q{\cal{D}}A~\exp\left\{-S[\bar{q},q,A]+\int
d^4x~\mu\bar{q}(x)\gamma_{4}q(x)\right\},
\end{eqnarray}
where $S[\bar{q},q,A]$ is the standard Euclidean QCD action with $q$
being the quark field with two flavors (we confine ourselves to the
two-flavor case with exact isospin symmetry and set
$\mu_u=\mu_d=\mu$, where $\mu_u$ and $\mu_d$ are the chemical
potential of the up and down quarks) and three colors. Here we leave
the ghost field term and its integration measure to be understood.
The pressure density ${\cal{P}}(\mu)$ is given by
\begin{equation}
{\cal{P}}(\mu)=\frac{1}{{\cal{V}}}~\ln\cal{Z}[\mu],
\end{equation}
where ${\cal{V}}$ is the four-volume normalizing factor. From this
one immediately obtains the quark-number density
\begin{eqnarray}
\rho(\mu)&=&\frac{\partial {\cal P}(\mu)}{\partial\mu}=\frac{1}{\cal{V}}\frac{1}{\cal{Z}[\mu]}\frac{\partial \cal{Z}[\mu]}{\partial\mu}\nonumber\\
&=&\frac{1}{{\cal{V}}}\frac{\int{\cal{D}}\bar{q}{\cal{D}}q{\cal{D}}A\int
d^4x
\bar{q}(x)\gamma_{4}q(x)\exp\left\{-S[\bar{q},q,A;\mu]\right\}}{\int{\cal{D}}\bar{q}{\cal{D}}q{\cal{D}}A~\exp\left\{-S[\bar{q},q,A;\mu]\right\}},
\end{eqnarray}
where $S[\bar{q},q,A;{\mu}]\equiv S[\bar{q},q,A]-\int d^4x~\mu
\bar{q}(x)\gamma_{4}q(x)$.

On the other hand, the full quark propagator at finite chemical
potential is defined as
\begin{equation}
G_{ij}[\mu](x,y)=\frac{\int{\cal{D}}\bar{q}{\cal{D}}q{\cal{D}}A~q_{i}(x)\bar{q}_{j}(y)\exp\left\{-S[\bar{q},q,A;\mu]\right\}}{\int{\cal{D}}\bar{q}{\cal{D}}q{\cal{D}}A~\exp\left\{-S[\bar{q},q,A;\mu])\right\}}.
\end{equation}
From Eq. (4) it is easy to obtain the following
\begin{equation}
\mathrm{Tr}\left\{G[\mu]\gamma_4\right\}=-\frac{\int{\cal{D}}\bar{q}{\cal{D}}q{\cal{D}}A\int
d^4x\bar{q}(x)\gamma_{4}q(x)\exp\left\{-S[\bar{q},q,A;\mu]\right\}}{\int{\cal{D}}\bar{q}{\cal{D}}q{\cal{D}}A~\exp\left\{-S[\bar{q},q,A;\mu]\right\}},
\end{equation}
where the notation $\mathrm{Tr}$ denotes trace over the color,
flavor, Dirac and coordinate space indices. Comparing Eq. (3) with
(5), we obtain a well-known result \cite{Bellac}
\begin{eqnarray}
\rho(\mu)&=&-\frac{1}{{\cal{V}}}\mathrm{Tr}\left\{G[\mu]\gamma_4\right\}=-N_cN_f
\int\frac{d^4p}{(2\pi)^4}\mathrm{tr}\left\{G[\mu](p)\gamma_4\right\},
\end{eqnarray}
where $N_c$ and $N_f$ denote the number of colors and of flavors,
respectively, and the trace operation $\mathrm{tr}$ is over Dirac
indices. From Eq. (6) it can be seen that the quark-number density
$\rho(\mu)$ is totally determined by the full quark propagator at
finite chemical potential. It should be noted that without loss of
generality, in deriving Eq. (6) we have used the unrenormalized
language. For the renormalized formulation  please see Ref.
\cite{Sun}.

The QNS is defined as the derivative of quark number density with
respect to the quark chemical potential
\begin{equation}
\chi=\frac{\partial \rho(\mu)}{\partial \mu}.
\end{equation}
Here, in order to show the difference between our approach and that
in the previous literatures, let us briefly recall the approach
in the previous literatures.  Substituting Eq. (6) into Eq. (7) and
adopting the identity
\begin{equation}
\frac{\partial G(p,\mu)}{\partial\mu}=(-)G(p,\mu)\frac{\partial
G^{-1}(p,\mu)}{\partial \mu}G(p,\mu),
\end{equation}
we have
\begin{equation}
\chi=N_cN_f\int\frac{d^4p}{(2\pi)^4}\mathrm{tr}[G(p,\mu)\frac{\partial
G^{-1}(p,\mu)}{\partial \mu}G(p,\mu)\gamma_4].
\end{equation}
Recall that the well-known Ward identity
\begin{equation}
i\Gamma_{\mu}(p,0)=\frac{\partial G^{-1}(p)}{\partial p_{\mu}},
\end{equation}
where $p$ denotes the relative momentum of the vector vertex and the
corresponding total momentum vanishes. Note that at finite chemical
potential, the fourth component of momentum
$\widetilde{p}_4=p_4+i\mu$. Then
\begin{equation}
(-)\Gamma_4(p,0;\mu)=\frac{\partial G^{-1}(p,\mu)}{\partial \mu}.
\end{equation}
Putting this equation into Eq. (9) and replacing the integration
over the fourth component of momentum with explicit summation over
Matsubara frequencies, we get the QNS at finite temperature and
chemical potential in the imaginary-time formalism
\begin{equation}
\chi(T,\mu)=(-)N_cN_fT\sum_{n=-\infty}^{+\infty}\int\frac{d^3p}{(2\pi)^3}\mathrm{tr}[G(\widetilde{p})\Gamma_4(\widetilde{p},0)G(\widetilde{p})\gamma_4],
\end{equation}
where $\widetilde{p}=(\overrightarrow{p},\omega_n+i\mu)$ with
fermion Matsubara frequencies $\omega_n=(2n+1)\pi T$. Thus a
model-independent closed integral formula is obtained, which
expresses the QNS in terms of the full quark propagator and the
vector vertex, both of the latter objects being basic quantities in
quantum field theory. From Eq. (12) it can be seen that in order to
calculate the QNS by means of this formula, one needs to know both
the full quark propagator and the vector vertex at finite $T$ and
$\mu$ in advance. For example, in previous literatures
\cite{Blaizot,Chakraborty,Blaizot1,Chakraborty1}, when those authors
calculate QNS in HTL approximation by means of Eq. (12), they have
to calculate the quark propagator and the quark-meson vertex in the
vector meson channel at zero total momentum separately. And they
have to further check whether the obtained quark-meson vertex and
quark propagator in HTL approximation satisfy the vector
Ward-Takahashi identity (as was pointed out in Ref.
\cite{Chakraborty1}, in a self-consistent calculation of QNS in HTL
approximation it is very important to ensure that the obtained
quark-meson vertex and quark propagator in HTL approximation satisfy
the vector Ward-Takahashi identity). In fact, one can use the vector
Ward-Takahashi identity to obtain the full vector quark-meson vertex
at zero total momentum directly from the full quark propagator (see,
for example, Ref. \cite{Zong}), so it is not necessary to calculate
the vector quark-meson vertex separately and check the validity of
the Ward-Takahashi identity. As will be shown below, with the aid of
the Ward-Takahashi identity it is not difficult to show that in the
calculation of QNS one only needs to know the full quark propagator
at finite $T$ and $\mu$.

Substituting Eq. (6) into Eq. (7) and replacing the integration over
the fourth component of momentum with explicit summation over
Matsubara frequencies, one can obtain a general formula for the QNS
at finite $T$ and $\mu$
\begin{eqnarray}
\chi(\mu,T)&=&-N_cN_f T\sum\limits_{n=-\infty}^{+
\infty}\int\frac{d^3\vec{p}}{(2\pi)^3}\frac{\partial}{\partial\mu}\mathrm{tr}\left[G[\mu,T](\vec{p},\omega_n)\gamma_4\right].
\end{eqnarray}
Here we note that at finite temperature and chemical potential, the
fourth component of momentum $\tilde{p_4}=\omega_n+i\mu$. Making use
of this fact, Eq. (13) can be rewritten as
\begin{eqnarray}
\chi(\mu,T)&=&-i N_cN_f T\sum\limits_{n=-\infty}^{+
\infty}\int\frac{d^3\vec{p}}{(2\pi)^3}\frac{\partial}{\partial
\tilde{p_4}}\mathrm{tr}\left[G[\mu,T](\vec{p},\omega_n)\gamma_4\right].
\end{eqnarray}
From this it can be seen that the QNS at finite chemical potential
and temperature is totally determined by the full quark propagator
$G[\mu,T](\vec{p},\omega_n)$. 
At this point some discussions are necessary. 
It should be noted that in the CJT (Cornwall-Jackiw-Tomboulis) or $\Phi$ functional approach, the grand canonical potential is not given by the propagator alone: an infinite series of irreducible diagrams is additionally needed. However, in some cases, the situation will be changed. For example, it can be shown that the nontrivial $\mu$ dependence of the partition function of QCD at finite chemical potential and zero temperature is totally determined by the full quark propagator at finite chemical potential and zero temperature (more details can be found in Ref. \cite{Sun}). As far as the present work, we do not claim that the partition function of QCD at both finite chemical potential and finite temperature is determined by the quark propagator alone. We just try to propose a direct and model-independent formula (Eqs. (13,14)) for calculating the QNS at finite chemical potential and temperature from the full quark propagator. In principle, once the exact form of $G[\mu,T](\vec{p},\omega_n)$ is known, one can have a thorough
understanding of the QNS at finite chemical potential and temperature. However, at present it is very difficult to calculate $G[\mu,T](\vec{p},\omega_n)$ from first principles of QCD. So when
one uses formula (13) or (14) to calculate $\chi(\mu,T)$, one has to resort to various nonperturbative QCD models. In this paper we shall use the HTL approximation to calculate the QNS at finite temperature and zero chemical potential by means of the above proposed direct method.  

It is generally believed that when the temperature $T$ is high
enough, the HTL approximation is a good approximation for QCD
\cite{Braaten,Braaten1,Braaten3,Blaizot2}. Based on this idea, the
authors of Ref. \cite{Blaizot} first calculated QNS in HTL
approximation. Later the calculation of QNS in HTL approximation was
further studied in Refs. \cite{Chakraborty,Blaizot1,Chakraborty1}.
Here we would like to stress that the purpose of our work is to give
a direct method to calculate the QNS. In this method the QNS is
totally determined by the full quark propagator at finite $T$ and
$\mu$. And in our work we choose the HTL quark propagator as an
example to illustrate how to apply this method to calculate the QNS
and compare the obtained result with the existing results of QNS in
the literature. Here we also would like to stress that in the
calculation of QNS in HTL approximation done in our work the vector
Ward-Takahashi identity is automatically satisfied.

In the HTL approximation, the quark propagator reads \cite{Braaten}:
\begin{equation}
\label{QP}
G(p)=-\frac{1}{D_+(p)}\frac{\gamma_4+i\hat{\textbf{p}}\cdot
\vec{\gamma}}{2}-\frac{1}{D_-(p)}\frac{\gamma_4-i\hat
{\textbf{p}}\cdot \vec{\gamma}}{2},
\end{equation}
where $p=(\vec{p},p_4)$, $\hat{\textbf{p}}=\vec{p}/|\vec{p}|$,
$p_4=(2n+1)\pi T ~(n \in \bf{Z})$ are the fermion Matsubara
frequencies and
\begin{eqnarray}
\label{D1}D_{\pm}(p)&=&-ip_4\pm |\vec{p}|+\frac{m_q^2}{|\vec{p}|}
\left[Q_0\left(\frac{ip_4}{|\vec{p}|}\right)\mp Q_1\left(\frac{ip_4}
{|\vec{p}|}\right)\right].
\end{eqnarray}
In Eq. (\ref{D1}), $m_q\equiv gT/\sqrt{6}$ is the quark ``thermal
mass'' with $g$ being the strong coupling constant.  $Q_0$ and $Q_1$
are Legendre functions of the second kind. Just as was shown above, 
we have two equivalent formulas (Eqs. (13) and (14)) for calculating the QNS
at finite chemical potential and temperature. In actual calculation of the QNS using which formula depends on which formula is more convenient. If one tries to use Eq. (13) to calculate the QNS at finite temperature and zero chemical potential in HTL approximation, one should use $\mu$-dependent HTL quark propagator (for very small values of $\mu$) and then take the limit $\mu \rightarrow 0$ after $\partial/\partial \mu$ operation in Eq. (13). It is generally believed that when $T$ or $\mu$ is high enough, the HTL or Hard-Dense-Loop (HDL) quark propagator is a good approximation for QCD. 
So far we do not have a reliable quark propagator at high $T$ and small $\mu$. Hence it is not convenient to calculate the QNS at finite temperature and zero chemical potential in HTL approximation using Eq. (13). However, in Eq. (14) the $\partial/\partial \mu$ operation has been replaced by $\partial/\partial \tilde{p}_4$ operation. So, if one tries to calculate the QNS at zero $\mu$ and finite $T$ in the HTL approximation, one can directly set $\mu=0$ in Eq. (14) and then input the HTL quark propagator at $\mu=0$ (Eq. (15)) into Eq. (14). By this method one obtains a formula for the QNS at zero $\mu$ and finite $T$ in the HTL
approximation:
\begin{eqnarray}
\label{Sus1}\chi(T)&=&-iN_cN_f T\sum\limits_{n}\int
\frac{d^3\vec{p}}{(2\pi)^3}\mbox{tr}\left(
\frac{\partial{G}}{\partial p_4}\gamma_4\right) =2iN_cN_f
T\sum\limits_{n}\int \frac{d^3\vec{p}}{(2\pi)^3}\left(\frac{1}{D_+}+
\frac{1}{D_-}\right)^{\prime}\nonumber\\
&=&-2iN_cN_f T\sum\limits_{n}\int
\frac{d^3\vec{p}}{(2\pi)^3}\left(\frac{D_+^{\prime}}{D_+^2}+
\frac{D_-^{\prime}}{D_-^2}\right)\label{sus3},
\end{eqnarray}
where $^{\prime}$ means $\partial/\partial p_4$. From Eq. (\ref{D1})
we can easily obtain the following
\begin{eqnarray}
\label{D2}D_+^{\prime}&=&-i+\frac{m_q^2}{|\vec{p}|}\left[
\left(1-\frac{ip_4}{|\vec{p}|}\right)Q_0^{\prime}-
\frac{i}{|\vec{p}|}Q_0\right],\\
\label{D3}D_-^{\prime}&=&-i+\frac{m_q^2}{|\vec{p}|}\left[
\left(1+\frac{ip_4}{|\vec{p}|}\right)Q_0^{\prime}+
\frac{i}{|\vec{p}|}Q_0\right].
\end{eqnarray}
To calculate the frequency sum in Eq. (\ref{Sus1}), let us consider
the following contour integral
\begin{eqnarray}
\int_{C_1\cup C_2}\frac{dp_4}{2\pi}\left(\frac{D_+^\prime}{D_+^2}+
\frac{D_-^{\prime}}{D_-^2}\right)\frac{1}{2}\tanh\left(
\frac{ip_4}{2T}\right),
\end{eqnarray}
where the integral contours $C_1$ and $C_2$ are shown in Fig. 1. The
contours are carefully selected because
$(D_+^\prime/D_+^2+D_-^\prime/D_-^2)$ has a branch cut from
$-i|\vec{p}|$ to $i|\vec{p}|$. When $g\rightarrow 0$, the contour
turns into the usually employed one which can be found in textbooks
of thermal field theory (see, for example, Ref. \cite{textbook}).

\begin{figure}
\centering \includegraphics[width=10cm]{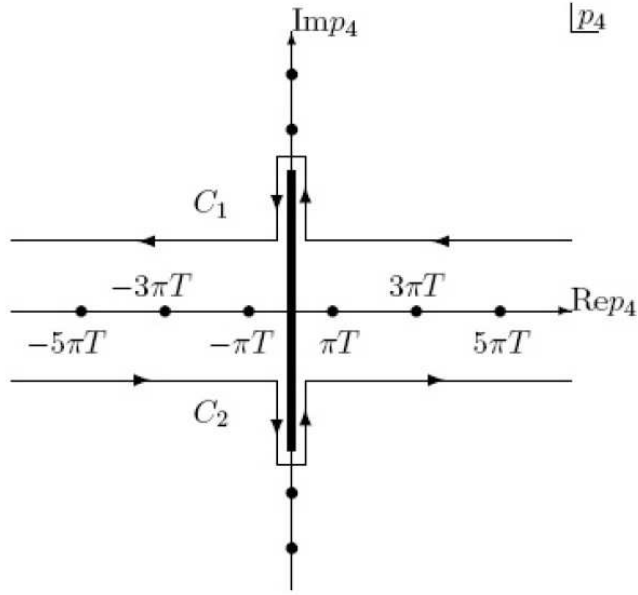} \vspace{-0.8cm}
\caption{The integral contours in complex $p_4$ plane}
\end{figure}

Applying Cauchy's theorem one has the following
\begin{eqnarray}
&&\int_{C_1\cup C_2}\frac{dp_4}{2\pi}\left(\frac{D_+^\prime}{D_+^2}+
\frac{D_-^{\prime}}{D_-^2}\right)\frac{1}{2}\tanh\left(
\frac{ip_4}{2T}\right)\nonumber\\
\label{int1}&=&T\sum_{n}\left(\frac{D_+^\prime}{D_+^2}+
\frac{D_-^{\prime}}{D_-^2}\right)+\int_{-i|\vec{p}|}
^{i|\vec{p}|}\frac{dp_4}{2\pi}\mbox{Disc}\left[\left(
\frac{D_+^\prime}{D_+^2}+
\frac{D_-^{\prime}}{D_-^2}\right)\frac{1}{2}\tanh\left(
\frac{ip_4}{2T}\right)\right].
\end{eqnarray}
To evaluate the integrals on the two sides of Eq. (21), we first use
integration by parts to obtain
\begin{equation}
\int_{C_1\cup C_2}\frac{dp_4}{2\pi}\left(\frac{D_+^\prime}{D_+^2}+
\frac{D_-^{\prime}}{D_-^2}\right)\frac{1}{2}\tanh\left(
\frac{ip_4}{2T}\right)=\int_{C_1\cup
C_2}\frac{dp_4}{2\pi}\left(\frac{1}{D_+}+\frac{1}{D_-}\right)
\frac{i}{T}\frac{\exp(ip_4/T)} {[\exp(ip_4/T)+1]^2} \label{int2}
\end{equation}
and
\begin{eqnarray}
&&\int_{-i|\vec{p}|}
^{i|\vec{p}|}\frac{dp_4}{2\pi}\mbox{Disc}\left[\left(
\frac{D_+^\prime}{D_+^2}+
\frac{D_-^{\prime}}{D_-^2}\right)\frac{1}{2}\tanh\left(
\frac{ip_4}{2T}\right)\right]\\
&=&\int\limits_{-i|\vec{p}|+0^+}
^{i|\vec{p}|+0^+}\frac{dp_4}{2\pi}\left[\left(
\frac{D_+^\prime}{D_+^2}+
\frac{D_-^{\prime}}{D_-^2}\right)\frac{1}{2}\tanh\left(
\frac{ip_4}{2T}\right)\right] -\int\limits_{-i|\vec{p}|+0^-}
^{i|\vec{p}|+0^-}\frac{dp_4}{2\pi}\left[\left(
\frac{D_+^\prime}{D_+^2}+
\frac{D_-^{\prime}}{D_-^2}\right)\frac{1}{2}\tanh\left(
\frac{ip_4}{2T}\right)\right]\nonumber\\
&=&\int\limits_{-i|\vec{p}|+0^+}
^{i|\vec{p}|+0^+}\frac{dp_4}{2\pi}\left(\frac{1}{D_+}+
\frac{1}{D_-}\right)\frac{i}{T}\frac{\exp(ip_4/T)}
{[\exp(ip_4/T)+1]^2} -\int\limits_{-i|\vec{p}|+0^-}
^{i|\vec{p}|+0^-}\frac{dp_4}{2\pi}\left(\frac{1}{D_+}+
\frac{1}{D_-}\right)\frac{i}{T}\frac{\exp(ip_4/T)}
{[\exp(ip_4/T)+1]^2}\label{int3}.\nonumber
\end{eqnarray}
Here we note that when Eq. (23) is obtained by integration by parts,
the contributions from end points cancel each other.

In the complex $p_4$-plane, the function
\begin{eqnarray}
\label{F1}g(p_4)&\equiv&\left(\frac{1}{D_+}+
\frac{1}{D_-}\right)\frac{\exp(ip_4/T)} {[\exp(ip_4/T)+1]^2}
\end{eqnarray}
has four poles and a branch cut from $-i|\vec{p}|$ to $i|\vec{p}|$
in the imaginary $p_4$ axis (See Fig. 1). These four poles are
located at $z_k=\pm i\omega_{\pm}(|\vec{p}|),k=1,\ldots,4$ and
$\omega_{\pm}(|\vec{p}|)$ $(\omega_{\pm}(|\vec{p}|)>|\vec{p}|)$ are
determined by the following equations
\begin{eqnarray}
\label{so1}\frac{|\vec{p}|(\omega_+-|\vec{p}|)}{m_q^2}-1&=&\frac{1}
{2}\left(1-\frac{\omega_+}{|\vec{p}|}\right)\ln\frac
{\omega_++|\vec{p}|}{\omega_+-|\vec{p}|},\\
\label{so2}\frac{|\vec{p}|(\omega_-+|\vec{p}|)}{m_q^2}+1&=&\frac{1}
{2}\left(1+\frac{\omega_-}{|\vec{p}|}\right)\ln\frac
{\omega_-+|\vec{p}|}{\omega_--|\vec{p}|}.
\end{eqnarray}
By closing the integral contour $C_1$ and $C_2$ by large
half-circles (note that the statistical factor is bounded on the
circle) we can get the following
\begin{eqnarray}
\frac{i}{T}\int_{C_1\cup C_2}
\frac{dp_4}{2\pi}g(p_4)&=&\frac{i}{2\pi T}\left\{-2\pi
i\sum\limits_{k=1}^4 \mbox{Res}[g(z_k)]\right\}.
\end{eqnarray}
Using
\begin{eqnarray}
\label{Pole1}&&\mbox{Res}\frac{1}{D_+(-
i\omega_+)}=\mbox{Res}\frac{1}{D_-(i\omega_+)}=i
\frac{\omega_+^2-|\vec{p}|^2}{2m_q^2}\equiv iZ_+(|\vec{p}|),\\
\label{Pole2}&&\mbox{Res}\frac{1}{D_+(
i\omega_-)}=\mbox{Res}\frac{1}{D_-(-i\omega_-)}=
i\frac{\omega_-^2-|\vec{p}|^2}{2m_q^2}\equiv iZ_-(|\vec{p}|),
\end{eqnarray}
the sum of residues in Eq. (27) is calculated to be
\begin{eqnarray}
\sum\limits_{k=1}^4
\mbox{Res}[g(z_k)]&=&2i\left\{\frac{Z_+\exp(\omega_+/T)}
{[\exp(\omega_+/T)+1]^2}+\frac{Z_-\exp(\omega_-/T)}
{[\exp(\omega_-/T)+1]^2}\right\}.
\end{eqnarray}
Therefore we obtain
\begin{eqnarray}
\frac{i}{T}\int_{C_1\cup C_2} \frac{dp_4}{2\pi}g(p_4)&=&\frac{2i}{T}
\left\{\frac{Z_+\exp(\omega_+/T)}
{[\exp(\omega_+/T)+1]^2}+\frac{Z_-\exp(\omega_-/T)}
{[\exp(\omega_-/T)+1]^2}\right\}.\label{sum1}
\end{eqnarray}
The integral in Eq. (23) can be calculated to be
\begin{eqnarray}
&&\int_{-i|\vec{p}|}
^{i|\vec{p}|}\frac{dp_4}{2\pi}\mbox{Disc}\left[\left(\frac{1}{D_+}+
\frac{1}{D_-}\right)\frac{i}{T}\frac{\exp(ip_4/T)}
{[\exp(ip_4/T)+1]^2}\right]\nonumber\\
&=&-\frac{i}{T}\int_{-|\vec{p}|}^ {|\vec{p}|}
\frac{d\omega}{2\pi}[\rho_+(\omega)+\rho_-(\omega)]
\frac{\exp(\omega/T)} {[\exp(\omega/T)+1]^2},\label{sum2}
\end{eqnarray}
where $\rho_\pm(\omega,|\vec{p}|)$ are the familiar spectral
functions of the quark propagator $(x\equiv\omega/|\vec{p}|)$
\cite{Braaten}:
\begin{eqnarray}
\rho_\pm(\omega,|\vec{p}|)&=&2\pi[Z_\pm(|\vec{p}|)\delta(\omega-
\omega_\pm(|\vec{p}|))+Z_\mp(|\vec{p}|)\delta(\omega+
\omega_\mp(|\vec{p}|))]\nonumber\\
&&+\frac{\pi m_q^2(1\mp
x)\theta(1-x^2)}{|\vec{p}|^3}\left\{\left[1\mp x+\frac{m_q^2}
{|\vec{p}|^2}\pm\frac{m_q^2}{2|\vec{p}|^2}(1\mp x)
\ln\frac{1+x}{1-x}\right]^2\right.\nonumber\\
&&\left.+\frac{\pi^2m_q^4}{4|\vec{p}|^4}(1\mp x)^2\right\}^{-1}.
\end{eqnarray}
Combining Eqs. (\ref{int1}), (\ref{sum1}), (\ref{sum2}), we obtain
the following
\begin{eqnarray}
T\sum_{n}\left(\frac{D_+^\prime}{D_+^2}+
\frac{D_-^{\prime}}{D_-^2}\right) &=&\frac{2i}{T}
\left\{\frac{Z_+\exp(\omega_+/T)}
{[\exp(\omega_+/T)+1]^2}+\frac{Z_-\exp(\omega_-/T)}
{[\exp(\omega_-/T)+1]^2}\right\}\nonumber\\
&&+\frac{i}{T}\int_{-|\vec{p}|}^ {|\vec{p}|}
\frac{d\omega}{2\pi}[\rho_+(\omega)+\rho_-(\omega)]
\frac{\exp(\omega/T)} {[\exp(\omega/T)+1]^2}\label{sum3}\nonumber\\
&=&\frac{2i}{T}[Z_+
n_F(\omega_+)(1-n_F(\omega_+))+Z_-n_F(\omega_-)(1-n_F(\omega_-))]\nonumber\\
&&+\frac{i}{T}\int_{-|\vec{p}|}^ {|\vec{p}|}
\frac{d\omega}{2\pi}[\rho_+(\omega)+\rho_-(\omega)]
n_F(\omega)(1-n_F(\omega))\label{sum4},
\end{eqnarray}
where $n_F$ is the Fermi distribution function
\begin{equation}
\label{Fermi1}n_F(\omega)\equiv \frac{1}{\exp(\omega/T)+1}.
\end{equation}
Substituting Eq. (\ref{sum4}) into Eq. (\ref{sus3}), we obtain the
QNS in HTL approximation
\begin{eqnarray}
\label{Sus4}&&\chi(T)=\chi_q(T)+\chi_L(T)\nonumber\\
&=&\frac{4N_cN_f}{T}\int
\frac{d^3\vec{p}}{(2\pi)^3}\left[\frac{\omega_+^2-\vec{p}^2}
{2m_q^2}n_F(\omega_+)(1-n_F(\omega_+)) +\frac{\omega_-^2-\vec{p}^2}
{2m_q^2}n_F(\omega_-)(1-n_F(\omega_-))\right]\nonumber\\
\label{Sus5}&&+\frac{2N_cN_f}{T}\int \frac{d^3\vec{p}}{(2\pi)^3}
\int_{-|\vec{p}|}^
{|\vec{p}|}\frac{d\omega}{2\pi}[\rho_+(\omega)+\rho_-(\omega)]
n_F(\omega)(1-n_F(\omega)),
\end{eqnarray}
where $\chi_q(T)$ is the pole-pole contributions from the
quasi-particle mode
\begin{equation}
\chi_q(T)=\frac{4N_cN_f}{T}\int
\frac{d^3\vec{p}}{(2\pi)^3}\left[\frac{\omega_+^2-\vec{p}^2}
{2m_q^2}n_F(\omega_+)(1-n_F(\omega_+)) +\frac{\omega_-^2-\vec{p}^2}
{2m_q^2}n_F(\omega_-)(1-n_F(\omega_-))\right]
\end{equation}
and $\chi_L(T)$ is the cut contributions from the Landau damping
\begin{eqnarray}
\chi_L(T)&=&\frac{2N_cN_f}{T}\int \frac{d^3\vec{p}}{(2\pi)^3}
\int_{-|\vec{p}|}^
{|\vec{p}|}\frac{d\omega}{2\pi}[\rho_+(\omega)+\rho_-(\omega)]
n_F(\omega)(1-n_F(\omega)).
\end{eqnarray}
It is obvious that the obtained QNS contains the contributions from
both the Landau damping and the quasi-particle poles. This result is
quite different from that of Ref. \cite{Chakraborty}. In Ref.
\cite{Chakraborty} the authors argue that ``the cut contributions
due to space-like quark momenta do not contribute because of the
number conservation''. In our opinion, although the number
conservation does hold, the quasi-particle mode and the Landau
damping mode are excited at different energy levels and therefore
have different statistical factors. As a result, the final result
contains both two contributions. Furthermore, even if one ignores the fact that the cut contribution from the Landau damping is nonzero, our result for the contribution from quasi-particle mode $\chi_q(T)$ still disagrees with the result of Ref. \cite{Chakraborty} (please compare Eq. (35) of Ref. \cite{Chakraborty} with Eq. (37) in the present work).
Here we should note
that various approximations employed in the literature lead to
different results (see. e.g., Ref. \cite{Blaizot1}), which shows the sensitivity
of the quantity under consideration and points to the need of a
concise approach to arrive at a unique result. This is one of our
motivations of doing this work.

The strong coupling constant $g(T)$ can be expressed as
\begin{eqnarray}
g(T)&=&\sqrt{\frac{4\pi\times12\pi}{(33-2N_f)\ln(Q^2/\Lambda_0^2)}},
\end{eqnarray}
where $\Lambda_0=200\sim300$MeV and $Q$ is the momentum scale. In
numerical calculations, we take $N_f=2$, $Q=2\pi T$ \cite{Braaten2}
and the phase transition point $T_c=0.49 \Lambda_0$ (about
$98 \sim 147$ MeV). 
\begin{figure}[h]
\centering \includegraphics[width=10cm]{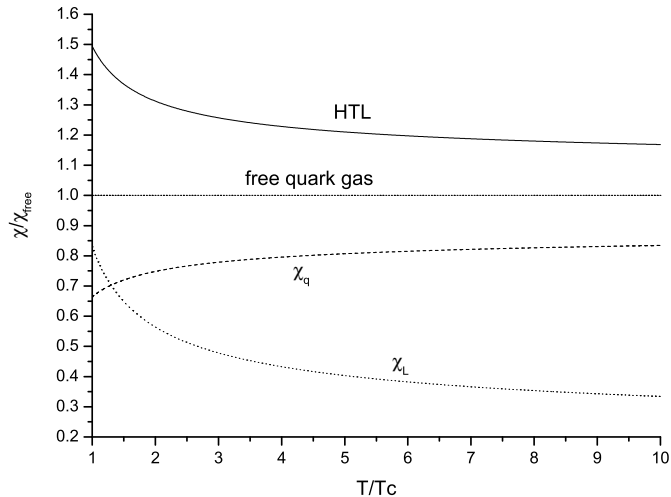} \vspace{-0.8cm}
\caption{$\chi/\chi_{free}$ as a function of $T/T_c$, where
$\chi_{free}=N_cN_fT^2/3$ is the QNS of the free quark gas}
\end{figure}

The numerical result is shown in Fig. 2. Just as is shown by Eqs. (36)-(38), the QNS has contributions from both quasi-particle mode
and Landau damping. In order to show clearly these two
contributions, in Fig. 2 we plot the contribution from the
quasi-particle mode and the contribution from Landau damping
separately. From Fig. 2 it can be seen that if one only considers
the contribution from quasi-particle mode, as was done in Ref.
\cite{Chakraborty}, then the QNS calculated in HTL approximation
will be smaller than the corresponding susceptibility of free quark
gas in the whole range of $T$, and when $T\rightarrow+\infty$ the
quasi-particle part will tend to the free quark gas result which can
be easily seen from Eq. (37) (when $T\rightarrow+\infty$,
$Z_+\rightarrow1$, $Z_-\rightarrow0$ and $\omega_\pm\rightarrow
|\vec{p}|$). Fig. 2 also shows that the contribution from Landau
damping is important even for $T\sim10T_C$. For the free quark gas,
there are only particle excitations, whereas in the calculation of
QNS in the HTL approximation done in this paper there are both
quasi-particle excitations and Landau damping mode excitations. It
is just the fact that the contribution of Landau damping is
important which renders the QNS calculated in HTL approximation
larger than the QNS of the free quark gas.

Here we should notice that the result in Fig. 2 is different from
the lattice result in Ref. \cite{Gavai}. It also differs from some
other existing results within the HTL approximation
\cite{Chakraborty,Blaizot2} and in the PNJL model \cite{Ghosh1,Ghosh2}. 
The origin of the difference between the result of QNS in Fig. 2 and those in the previous literatures can be understood as follows. From Eq. (17) it can be seen that in order to calculate the QNS in HTL approximation, one needs to know the HTL quark propagator in the whole momentum range. It is well-known that the HTL approximation is only valid for external momentum much smaller than $T$. If one ignores this fact and assumes that the HTL quark propagator (15) is applicable in the whole momentum range (this is in the same spirit as the one for calculating QNS in HTL approximation \cite{Chakraborty,Chakraborty1}), then one will obtain the QNS result shown in Fig. 2. Therefore the result in Fig. 2 is a consequence of ignoring the the range of applicability of the HTL quark propagator (15). So in a consistent calculation of QNS in the HTL approximation by means of Eq. (17) one should introduce a momentum cutoff $\Lambda_{HTL}$ below which the HTL quark propagator (15) is applicable. In our calculation we choose $\Lambda_{HTL}=gT$. We make this choice because $gT$ is an important energy scale to identify the soft momentum in HTL approximation. If we limit the range of integration of the contribution of Landau damping in Eq. (38) to the region $|{\vec p}|\leq gT$, we will get the numerical result of the QNS shown in Fig.3. From Fig. 3 it can be seen that the QNS under HTL approximation is smaller than the QNS of free quark gas in the whole range of $T$ and the contribution of the quasi-particle mode is much larger than that of Landau damping. Here we should
point out that the result in Fig. 3 is in fact not sensitive to the
variation of the cut-off $gT$. For example if one set the cut-off to
be $2gT$ one would find the change of $\chi$ is about $10\%$. This
is because setting cut-off in the order of $gT$ would result in $\chi_L <<
\chi_q$, and consequently, the quark number susceptibility is
determined essentially within the given approximations by the quark
quasi-particle contribution. Therefore one would find that the result of
$\chi(T)$ shown in Fig. 3 is in agreement with the results in Refs.
\cite{Gavai,Blaizot2} on the level of $10$ percents. 
In addition, it should be noted that our results are gauge independent due to gauge invariance of the HTL quark propagator.
We also want to point out that in the calculation of QNS in the present paper we have used HTL quark propagator at one-loop order. In fact, the method employed in the present paper can be easily applied to the calculation of the QNS using HTL quark propagator up to two-loop or higher order.
\begin{figure}[h]
\centering \includegraphics[width=10cm]{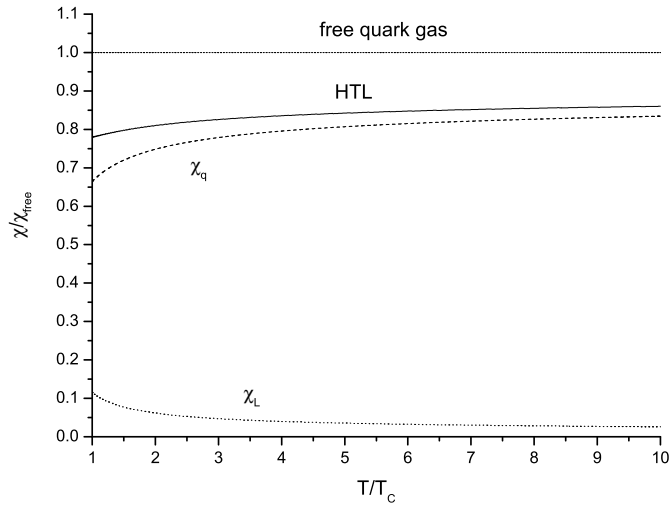} \vspace{-0.8cm}
\caption{$\chi/\chi_{free}$ with range of integration of contribution of Landau damping limited in the
region $p<gT$}
\end{figure}

To summarize, in the present work we derive a new and model-independent formula (Eq. (14))
for calculating QNS, which expresses the QNS totally in terms of the full quark propagator at finite temperature and chemical potential. Then we choose the HTL quark propagator at zero chemical potential (Eq. (15)) as an approximation to the
full quark propagator to calculate the QNS at finite temperature and zero chemical potential by means of our method (see Eq. (17) in our paper). As is shown in Eq. (17), when one calculates the QNS
at finite $T$ and zero $\mu$, one should integrate the momentum from zero to infinity. 
However, it is well-known that HTL quark propagator is only applicable for external momentum much smaller than the temperature $T$ and one should choose a boundary value $\Lambda_{HTL}$
below which the HTL quark propagator is applicable. In this paper we have chosen $\Lambda_{HTL}=gT$ 
because $gT$ is an important energy scale to identify the soft momentum in HTL approximation.
When limiting the range of integration of the contribution of Landau damping to be $p<gT$,
we find that the result of QNS calculated by our method is consistent with the results of previous literatures.

\bigskip
\begin{acknowledgements}
{\noindent \bf \large Acknowledgements}\\
We thank J.-P. Blaizot for discussions on HTL approximation. This work is supported in part by the National Natural Science Foundation of China (Grant Nos. 10775069 and 10935001) and the Research Fund
for the Doctoral Program of Higher Education (Grant Nos. 20060284020 and 200802840009).
\end{acknowledgements}

\end{document}